\documentclass{sig-alternate-05-2015}
\usepackage{algorithm2e}
\usepackage{comment}
\usepackage{url}
\pdfoutput=1
\begin{document}

\setcopyright{acmcopyright}





%

\title{Investigating Factors Influencing the Latency of  Cyberbullying Detection}
%
%
%
%
%

\numberofauthors{5} 
%
\author{
%
%
\alignauthor
Rahat Ibn Rafiq\\
       \affaddr{University of Colorado Boulder}\\
       \affaddr{Boulder, CO, USA}\\
       \affaddr{Rahat.Rafiq@Colorado.Edu}\\
\alignauthor 
Homa Hosseinmardi\\
       \affaddr{University of Colorado Boulder}\\
       \affaddr{Boulder, CO, USA}\\
       \affaddr{Homa.Hosseinmardi@Colorado.Edu}\\
\alignauthor
Richard Han\\
       \affaddr{University of Colorado Boulder}\\
       \affaddr{Boulder, CO, USA}\\
       \affaddr{rhan@cs.colorado.edu}\\
\and  
\alignauthor 
Qin Lv\\
       \affaddr{University of Colorado Boulder}\\
       \affaddr{Boulder, CO, USA}\\
       \affaddr{Qin.Lv@Colorado.Edu}\\
\alignauthor
 Shivakant Mishra\\
       \affaddr{University of Colorado Boulder}\\
       \affaddr{Boulder, CO, USA}\\
       \affaddr{mishras@cs.colorado.edu}\\
}

\maketitle

\begin{abstract}
Cyberbullying in online social networks has become a critical problem,
especially among teenagers who are social networks' prolific users.
As a result, researchers have focused on identifying distinguishing features
of cyberbullying and developing techniques to automatically detect
cyberbullying incidents. While this research has resulted
in developing highly accurate classifiers, two key practical issues related
to identifying cyberbullying have largely been ignored, namely scalability of
cyberbullying detection services and timeliness of raising alerts whenever
a cyberbullying incident is suspected. 

These two issues are the
subject of this paper. We propose a multi-stage cyberbullying detection
solution that drastically reduces the classification time and the time
to raise cyberbullying alerts. The proposed solution is highly scalable, does
not sacrifice accuracy for scalability, and is highly responsive in raising
alerts. The solution is comprised of three novel components, an initial
predictor, a multilevel priority scheduler, and an incremental classification
mechanism. We have implemented this solution and utilized data obtained from the Vine
online social network to demonstrate the utility of each of these components 
via a detailed performance evaluation.  We show that our complete solution is 
significantly more scalable and responsive than the current state-of-the-art.

\end{abstract}

%
%





\section{Introduction}
\label{sec:introduction}
Unprecedented growth in the popularity of online social networks (OSNs),
especially among teenagers, has unfortunately resulted in significant increase
in cyberbullying perpetrated via these networks. It has been reported that in the United States alone,
more than fifty percent of teenage OSN users have been affected by the threat of cyberbullying~\cite{halfOfTeens}. Devastating consequences of cyberbullying such as suicides \cite{Pheobe,hannah,askfmsuicides} have led researches to
detect cyberbullying incidents in OSNs like Ask.fm, Instagram, Vine etc \cite{rahatVineASONAM2015,homaSocinfoInstagram,asonam14,homa-cyberbullying-socialcom14}. These researches 
mostly followed tghe methodology of collecting data from OSNs, Label those data for cyberbullying and then
design a classifier to detect and predict cyberbullying incidents. These researches have also investigated 
the issue of identifying imbalance of power between perpetrators and victims,
which is a key feature of bullying, and distinguishing between
cyber-aggression and cyberbullying ~\cite{kowalski2012cyberbullying}, thus paving the way for 
highly accurate classifiers.

While highly accurate cyberbullying detection classifiers are undoubtedly
needed, there are two key practical issues that have been largely ignored
so far. The first issue concerns the scalability of the cyberbullying 
detection solutions. OSNs, of course, involve an enormous amount of data, in the
order of several hundred gigabytes per day. For example, it has been
reported that Vine, around $8,233$ videos are
shared every minute while the number of Vine loops played daily is $1.5$ billion~\cite{VineStatistics}.

The second issue concerns the timeliness of raising alerts whenever
cyberbullying incidents are suspected. Cyberbullying is different from
traditional, face-to-face bullying, because it can occur 24/7, perpetrators
can stay anonymous, and they have easy access to sophisticated tools to
launch cyberbullying attacks. Furthermore, the consequences of cyberbullying
can be disastrous and it is extremely important to provide the necessary support
to the victims as early as possible. As a result, a timely detection of
cyberbullying is of paramount importance, so that an alert can be raised as soon as possible.

In this paper, we propose a multi-stage cyberbullying detection solution 
designed to improve the scalability of cyberbullying detection as well as
reduce the time to raise an alert. \emph{To the best of 
our knowledge, we are the first to propose a scalable and responsive solution to cyberbullying
detection in OSNs}. A key property of the solution is that it 
achieves sufficient classification accuracy while accomplishing these two goals. The
solution consists of three key components, namely an initial prediction stage for
prioritization and scaling, a dynamic, multilevel priority scheduler for
improved responsiveness, and an incremental feature extraction and
classification stage for further scaling. We have built a prototype of
this solution. Using online social networking data from Vine, we demonstrate
the utility of each of these components, and show that our complete cyberbullying detection solution
is significantly more scalable and responsive than the current
state-of-the-art. We make the following important contributions:

\begin{itemize}
\item We propose an incremental computational design for feature extraction and classification
reusing previous classification results reducing the time-overhead with minimal impact on the accuracy
\item We propose a dynamic, multi-level priority scheduler that assigns high preference to
potential cyberbullying media-sessions, thereby improving responsiveness of the solution
\item We propose an initial predictor that only uses a small amount of session data to quickly 
assign a fairly accurate initial priority for the dynamic priority scheduler
\item Finally, using real world data from Vine, we demonstrate the utility of each stage of our solution as
well as the computational scalability and responsiveness of the whole solution 
\end{itemize}

\section{Related Work}

As mentioned earlier, the majority of research on cyberbullying detection has focused on improving the accuracy of cyebrbullying detection classifiers\cite{Ptaszynski2010,usingML,Nahar13,dinakar2011modeling,homaSocinfoInstagram,rahatVineASONAM2015}. Several cyberbullying detection applications have been developed in recent years \cite{cyberbullyingApps}. Most of these applications (e.g., Mobicip) introduce parental control including category blocking, time limits, Internet activity reports, blocked phrases, and YouTube filtering\cite{cyberbullyingApps} whereas others (e.g., iAnon and GoGoStat) search for specific profanity words\cite{cyberbullyingApps}. As shown in \cite{instagramCyberbullyingDetection,homaSocinfoInstagram}, profanity can not solely be an indicator of cyberbullying. So it is important to make use of the definition of cyberbullying and take into account the repetitive nature of perpetrated aggressiveness and imbalance of power while building an accurate cyberbullying detection solution for online social networks.

As far as we know, none of the prior research has addressed the issue of computational
scalability and/or responsiveness in the context of cyberbullying detection. The issue of scalability has been, however, an important factor for other research areas such as misbehavior detection in online video chat services \cite{ScalabaleVideoChat,efficientmisbehaviorwsdm2012} ,cyber-attack detection in communities \cite{ScalableCyberAttacks} and online advertising \cite{scalableAdvertisingwsdm2014}.

\section{Problem Definition and Data Set}
\label{sec:data_collection}

\begin{figure*}
\centering
\includegraphics[width=0.8\textwidth, height=0.25\textwidth]{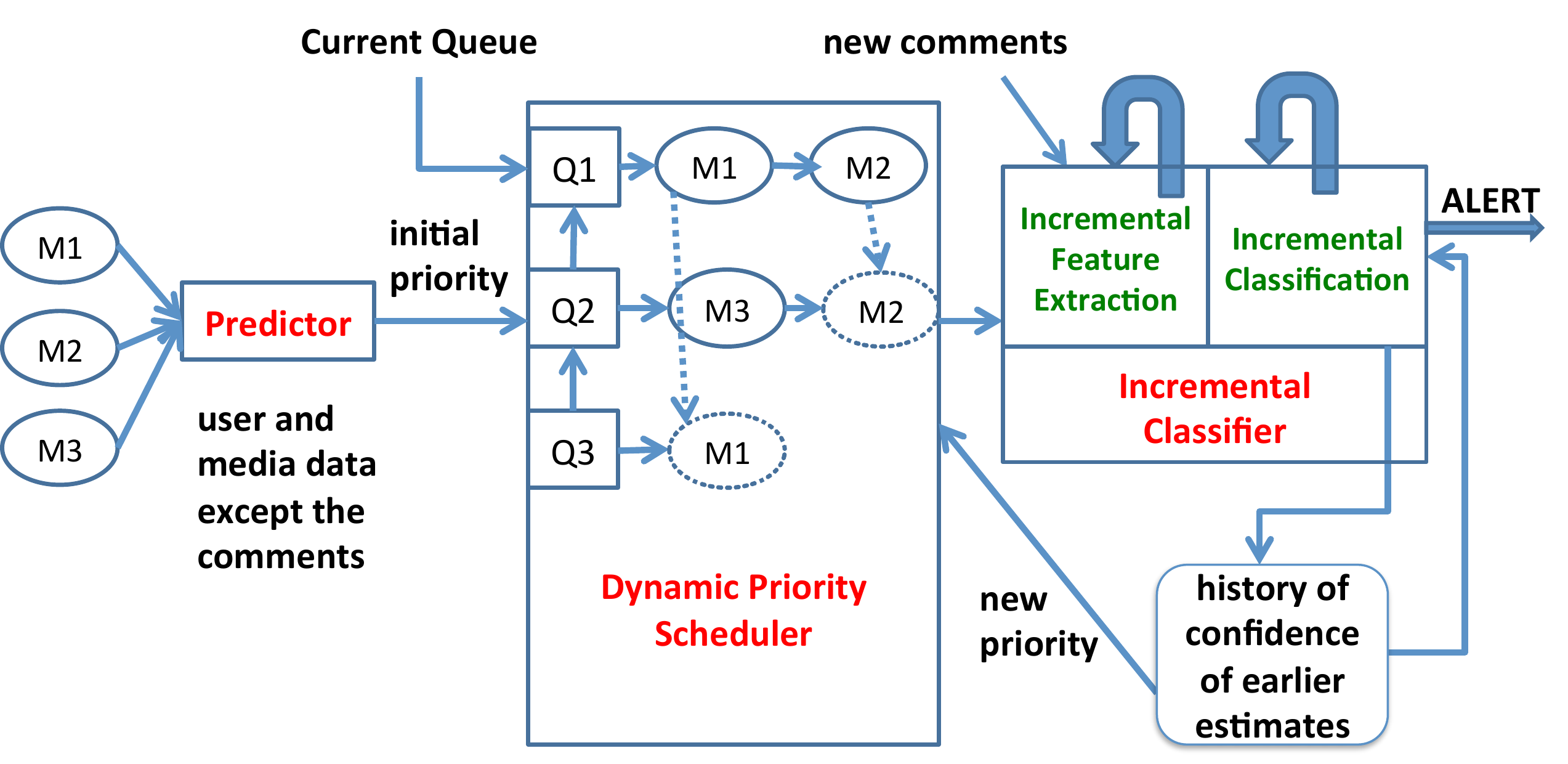}
\vspace{-5mm}
\caption{Scalable and responsive cyberbullying detection architecture }
\label{fig:design_architecture}
\vspace{-5mm}
\end{figure*}

Our cyberbullying detection solution aims to determine whether a media 
session in an OSN consists of cyberbullying activities or not, where a media session 
is comprised of the media and all it's associated comments. All the posting 
activities of a media session are ordered in time, so earlier detection using  
a partial set of the  media session allows for more timely alerts of suspected cyberbullying. 

In this work, we use Vine data from ~\cite{rahatVineASONAM2015} that collected more than $100000$ media sessions. In ~\cite{rahatVineASONAM2015}, they used CrowdFlower, a crowd-sourcing platform, to label $983$ Vine media sessions as instances of cyberbullying or not. This labeled data-set is used as the ground truth for our cyberbullying detection solution.

\section{Design Overview}

Figure \ref{fig:design_architecture} illustrates the three components of our proposed solution: predictor,
dynamic priority scheduler and incremental classifier, which in turn contains incremental feature extraction
and incremental classification. A novel feature of this solution is that it assigns priorities to media session
based on how likely that session will be a cyberbullying one. High priority media sessions are examined earlier and more frequently than the rest based on the intuition that this will enable the solution to detect potential cyberbullying sessions much earlier, thereby improving the responsiveness of the solution.

Newly created media sessions are assigned an initial priority by the predictor component and then are placed in
the appropriate priority queue. The dynamic priority scheduler passes highest priority media session to the 
incremental classifier component. The incremental feature extraction updates the previously stored feature values
by processing and combining the newly arrived comments' feature values. The incremental classification then 
classifies the media session with certain confidence/probability. This confidence is then used to raise an alert
and dynamically change the priority of the media session to enable previously low priority but currently
high probability cyberbullying media sessions to attain high priority. We now describe the details of each of these three components.

\subsection{Initial Priority Prediction}

In Figure \ref{fig:design_architecture}, M1,M2 and M3 represent newly crated media sessions, which would be retrieved 
from Vine in real time when the solution is actually deployed. The initial predictor predicts the priority of a 
new media session as higher or lower depending on the later likelihood of that media session being cyberbullying.
The predictor essentially is a classifier that makes use of features from the user who posted the video of that
media session (number of followers, number of followings and number of media sessions shared by the user) and from 
the media session caption which is available when the vine video is first posted. Note that the predictor does not extract comment features which are not available when the video is first posted. The predictor runs once for each media session and 
assigns a high or low priority and then puts it in the dynamic priority scheduler. Because the predictor is the first component of the solution, it is imperative that it has a very high recall to ensure it does not fail to assign high priority to a 
potential cyberbullying media session. With the help of a very fast predictor, we can filter away highly unlikely cyberbullying media sessions and postpone their processing to a later time, so that high priority media sessions can be processed earlier.

\subsection{Dynamic Priority Scheduler}
The dynamic priority scheduler schedules media sessions for processing
based on their priorities. As shown in Figure~\ref{fig:design_architecture},
three separate queues, Q1, Q2 and Q3 are maintained. The scheduler schedules
media sessions in queue Q1 (pointed to by current head) for processing 
one by one in the
queue order. After a media session has been processed by the cyberbullying
detector component and if no alert is raised, it is placed at the end of
either queue Q2 or queue Q3 depending on the new priority assigned to it
(discussed in the next subsection).
If the session's new priority is high, it is placed in queue Q2, and if
the session's new priority is low, it is placed in queue Q3. When all
media sessions in queue Q1 have been processed, queue Q2 becomes queue Q1,
queue Q3 becomes queue Q2, and queue Q3 becomes empty.
In the example shown in Figure \ref{fig:design_architecture}, after the
initial prediction of the three media sessions M1, M2 and M3, M1 and M2 
are classified as higher priority and are initially place in queue Q1, while 
M3 is classified as lower priority and is placed in queue Q2. M1 is scheduled
first and is processed by the cyberbullying detector component. After this
processing, it is assigned a low priority, and so is added at the end of
queue Q3. M2 is scheduled next and is processed next.
After this processing, it is assigned a high priority, and so is added at
the end of queue Q2. At this time queue Q1 is empty, and queue Q2 becomes
queue Q1 and queue Q3 becomes queue Q2. As a result, M3 is scheduled next
followed by M2. This process then continues.

Our scheduler has three interesting properties. First, it dynamically changes the priority of a media
session based on its current status. As a result, a media session may have a high priority at one time
and low priority at some other time. Second, as will be shown later in our evaluation, having only two
levels of priority in the scheduler proves to be sufficient to improve responsiveness of our solution. Finally,
this dynamic scheduler ensures that no media session will starve, meaning every media session regardless of its priority 
is guaranteed to be processed by our solution at some future time.

\subsection{Incremental Classifier}
\label{subsec:incremental_classifier}

This component receives a media session from the scheduler, extracts features from newly arrived comments, updates 
the feature values for that media session by combining the previsouly stored feature values and classifies whether 
the current status of that media session constitute cyberbullying. To improve the scalability of the two sub-components
of feature extraction and classification, we employed incremental computation, in which we reuse previous results to reduce
the additional computation needed as new comments arrive.

In particular, we consider number of negative comments,number of negative words and summation of individual comment-text sentiments. These features by nature can be incrementally linear in the sense that once the values corresponding to these features have been computed for the first $n$ comments, then when $\delta n$ new comments arrive, we only have to compute the individual feature vector values for the new $\delta n$ comments while reusing the previous feature vector values for the previous $n$ comments to generate the overall feature vector values for the $n+\delta n$ comments. This dramatically reduces
the amount of computation because this approach is driven by  $\delta n$ at each invocation of this component instead of $n+\delta n$.

Similarly, the classification sub-component also seeks to employ incremental computation to improve its scalability.  We sought classifier algorithms that not only yield high accuracy, precision and recall, but also are capable of reusing previous classification results in order to reduce computation time.  We found that logistic regression (LR) was the most promising algorithm that met all of these goals.  The way LR works is as follows: if we have $n$ features $a_i,i=0,1,2,3...n$ and LR, after being trained, assigns an weight $w_i,i=0,1,2...n$ to each of those features, then LR computes the combined features value $c = \sum_{0}^{n}a_iw_i$. This value is then employed in a sigmoid function to output a value from 0 to 1, thus interpretable as a probability/confidence \cite{logisticRegression},( which we use for our dynamic scheduler, as explained later). Because of the nature of the linear function that LR uses, it is capable of incremental classification. For example, for a particular feature $s$, if the previously stored feature value for $n$ comments was $a_{sn}$ and after the newly arrived $\delta n$ comments, the new feature value is $a_{s(n+\delta n)}$, we only apply the new feature value to the linear function of LR if $a_{s(n+\delta n} \neq a_{sn}$, thus saving some computational effort.

By taking into account the classification history of a media session, an alert is sent to the appropriate authorities.
Alerts can be sent more than once for one media session because cyberbullying may was and wane within a media session, i.e., 
a media session for which an alert has been sent two days ago for comments until that point may also experience cyberbullying
for comment threads that may come several days later, for which our solution will be able to send alerts too.

Each invocation of our classifier also generates a confidence value indicating how confident it is about the 
decision it has made.\emph{ We use this confidence level to adjust the priority of the media session for future processing}.
Media session in which the classifier has a high and low confidence for cyberbullying are given higher and lower priority
respectively, thus enabling dynamic promotion or demotion of a media session's priority. \emph{ For example}, a low priority media session, after being classified by the classifier as not-cyberbullying with a confidence of $0.55$ might be assigned as high priority before being inserted in the dynamic scheduler because of the fact that it has a comparatively high likelihood of being a cyberbullying session in later phase due to the closeness of probabilities for cyberbullying and not-cyberbullying decisions ($0.45$ and $0.55$ respectively).

\section{Initial Predictor}

\begin{table}
\centering
\caption{Predictor Performance using Different Classifiers for Cyberbullying 
Detection}
\label{predictor_performance}
\begin{tabular}{|c||cc|} \hline
 Classifier      &  Precision  &  Recall  \\ \hline \hline 
 kNN  &0.43&0.82 \\ \hline
 Random Forest&0.43&0.66\\ \hline
 Decision Tree&0.42&0.61\\ \hline
 Naive Bayes&0.48&0.51\\ \hline
\textbf{Logistic Regression} & \textbf{0.44}      & \textbf{0.93}     \\ \hline
 \end{tabular}
 \vspace{-5mm}
\end{table}
In this section we investigate a classifier for the initial priority prediction when a new media session comes in. The purpose of this initial predictor is to run once, for each new media session, to provide a quick first estimate of whether that session should be subjected to high priority scrutiny by the solution.  We are able to make an initial estimate based only on the initial vine posting as well as features obtained from the posting user.

As noted earlier, the predictor should achieve both high recall and high 
efficiency (i.e., fast prediction). 
We chose features that could be quickly extracted, and a classifier that could be executed quickly at this stage. Features that could be easily calculated include user features (number of followers, number of followings and number of posts shared by that user), and media-session features  based on the media caption of initial post.
For example, sentiment analysis features such as the polarity and subjectivity sentiment of the media caption can be quickly obtained by employing Python's NLTK library \cite{textblob}. The library gives as output polarity and subjectivity value of a particular text. Texts have a polarity (negative/positive, -1.0 to +1.0) and a subjectivity (objective/subjective, +0.0 to +1.0) showing how negative and subjective a particular text is. The library is reported to have an accuracy of 75 percent \cite{textblob} when applied to a English movie review data-set \cite{movie-review}. This convinced us to use this library to extract sentiments  from the texts when designing features. We also employed other features too (number of times the media has been viewed, media caption unigram and so on) but our experiments found that by using the five features (number of followers, number of followings and number of posts shared by that user, media caption polarity and subjectivity), we gained the best results in terms of precision and recall.

Table \ref{predictor_performance} shows different classifiers' performance as a candidate for our predictor. We used $10$-fold cross validation on the labeled Vine dataset of $983$ media-sessions\cite{rahatVineASONAM2015} and measured only the cyberbullying class' precision and recall because we wanted to see which classifier candidate misses the least number of cyberbullying classes (i.e., high 
recall). It is worth mentioning that we only present the results of the classifiers with the best performing results. As is evident from the table, Logistic Regression gained the highest recall for the cyberbullying class, which means it assigned high priority to $93$ percent of the true cyberbullying media sessions.

To evaluate the speed of our predictor, 
we measured the time it takes to extract the features and make the prediction for different numbers of media sessions and found that for $1000$ media sessions the predictor takes less than a second to predict the priority. To put this into perspective, 
Vine receives $8233$ media sessions every minute \cite{VineStatistics}. Thus, from the apparent linear relationship between number of media sessions and the predictor's running time, we can estimate that to process those new $8233$ media sessions from Vine every minute, the predictor will take less than $8$ seconds.

\section{Incremental Classifier}

\subsection{Standard Classifiers}

\begin{table}
\centering
\caption{Comparison of Potential Standard Classifiers using the 983 Labeled Media Sessions}
\label{table:baseline_classifier_compare_table}
\begin{tabular}{|p{2.5cm}|l|l|l|} \hline
Classifier  & Precision & Recall & Run Time (s) \\ \hline \hline 
AdaBoost  & 0.7138              & 0.54      & 228    \\ \hline
Logistic Regression  &  0.71              & 0.66         & 44.42    \\ \hline
\end{tabular}
\vspace{-5mm}
\end{table}

When choosing the right classifier for cyberbullying detection, all prior 
research has focused on quality measures such as precision and recall. 
Since we are now addressing the critical issue of scalability, we must 
consider the running time of the classifier as well. 
In \cite{rahatVineASONAM2015}, the AdaBoost classifier was reported to have the best performance based on accuracy, precision and recall values, closely followed by the logistic regression classifier. Table ~\ref{table:baseline_classifier_compare_table} compares these two 
classifiers in terms of precision, recall, and running time. The features AdaBoost classifier used were number of followers and followings, likes and views for media sessions, media caption polarity and subjectivity, total number of negative comments, summation of negative comment polarity and subjectivity, total individual comment polarity, total individual comment subjectivity, total negative words, and total number of negative comments, unigrams based on comments. For Logistic Regression, the features used were number of followers, followings, media caption polarity and subjectivity,total individual comment polarity, total individual comment subjectivity, total negative words, and total number of negative comments. The 
performance values showed in Table \ref{table:baseline_classifier_compare_table} were obtained  
using $10$-fold cross validation on the $983$ labeled Vine media sessions collected from \cite{rahatVineASONAM2015}.
We notice that although the Adaboost classifier 
achieves a slightly higher precision, logistic regression achieves higher recall. Furthermore, the running time of logistic regression classifier is
significantly less, more than five times faster than that of the Adaboost classifier.
The reason for this is twofold. First, Adaboost needed unigram features to achieve a high precision and recall, but unigram feature extraction is computationally intensive.  In comparison, logistic regression is able to achieve effectively the same precision and better recall while using features that are much more lightweight to compute, thus yielding much lower running time. Second, Adaboost classifier is a meta-estimator that begins by fitting a classifier on the original data-set and then fits additional copies of the classifier on the same data-set but where the weights of incorrectly classified instances are adjusted such that subsequent classifiers focus more on difficult cases \cite{adaboost}, thus making it computationally extensive compared to much simpler logistic regression. Based on 
these analyses, we have chosen the logistic regression classifier for detecting 
cyberbullying in our solution. It is worth mentioning that we have employed other classifiers based on different combinations of features too (Decision Tree, Random Forest, Naive Bayes, Perceptron etc) and only present the classifiers and feature combinations that yielded the best results.

\subsection{Design of Incremental Classifier} 

We apply incremental computation to both the feature extraction and the logistic regression classifier to improve scalability. Based on prior work on cyberbullying
detection in Vine~\cite{rahatVineASONAM2015}, the features we consider 
for classification include summation of polarity and subjectivity of all individual comments belonging to a media session, number of total negative words, and number of total negative comments, where a comment is considered negative if it contains at least one negative word. We employ a negative word list provided in \cite{NegativeWordsList} as our negative word dictionary.  As noted in Section \ref{subsec:incremental_classifier}, these features are amenable to incremental feature extraction, enabling reuse of previous computational results to minimize computation of new features as new comments arrive.  Algorithm \ref{algo:incremental_feature_extraction} provides a pseudo-code of the incremental feature extraction algorithm for our incremental detector.

\begin{figure}
\vspace{-2mm}
\centering
\includegraphics[width=0.4\textwidth, height=0.25\textwidth]{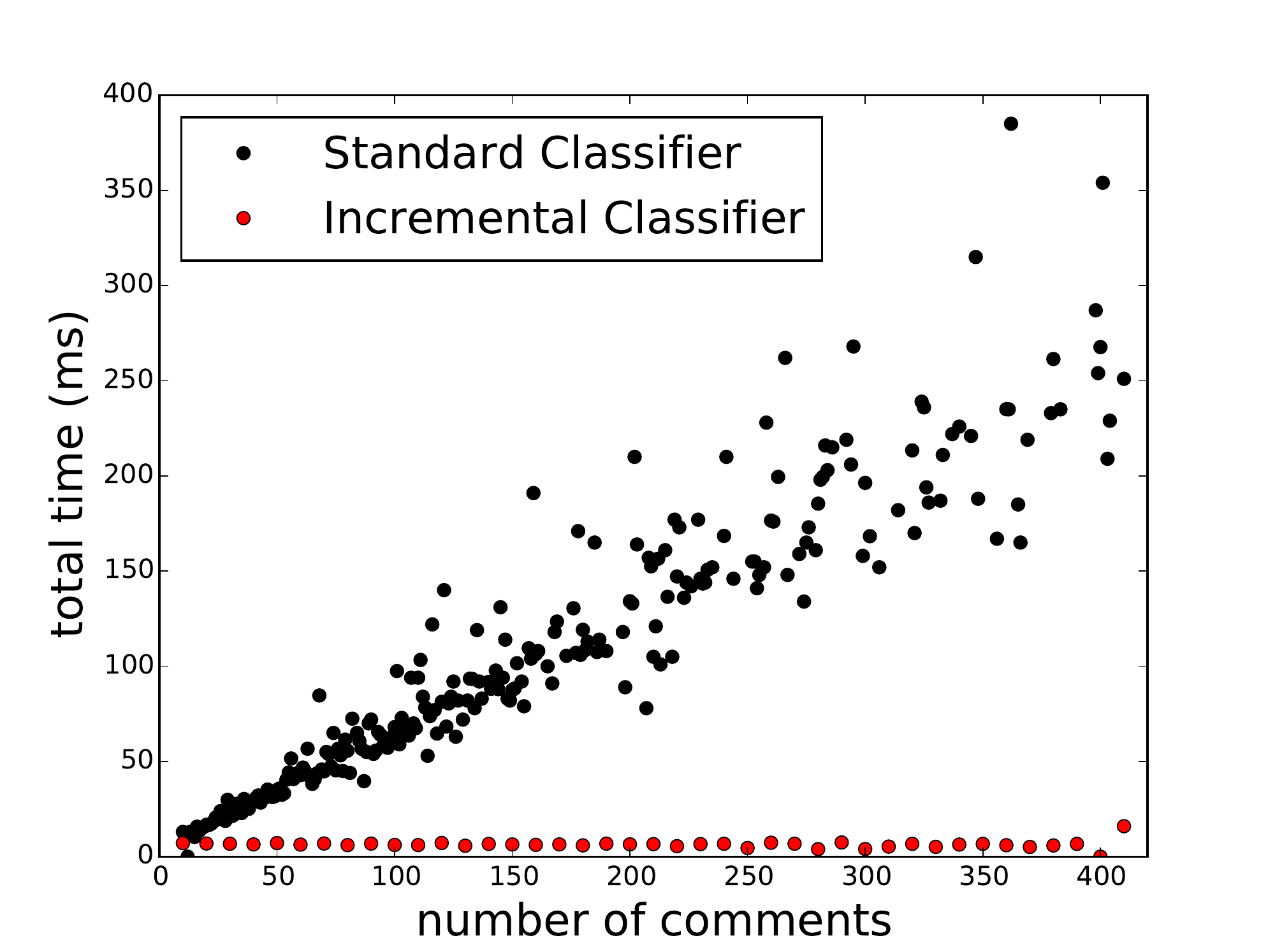}
\vspace{-5mm}
\caption{Total time taken by standard and incremental classifiers as new comments come in per media session.}
\vspace{-5mm}
\label{fig:average_time_per_media_session}
\vspace{-1mm}
\end{figure}

We also augment the standard logistic regression classifier with incremental computation.  Logistic regression takes as input a set of features $X$, and during the training process, the classifier generates a set of weights $\theta$  corresponding to those features. When a new media session comes in, the feature extraction step computes a matrix $X$ for that particular media session and computes $C=X.\theta$, which is then used to make the corresponding prediction. For the incremental feature extraction sub-component, we save the  $X_{old}$ value for the previous $n$ comments, compute $X_{\delta n}$ for the new set of $\delta n$ arrived comments and compute the new $X$ by combining $X_{old}$ and $X_{\delta n}$ instead of computing $X$ all over again for all $n + \delta n$ comments. For the incremental classification part, we only use those components of $X$ that have been changed to compute $C=X.\theta$ instead of doing the full $X.\theta$ computation. For this purpose, we save $\forall i, X_i.\theta_i$ where $X_i$ is the $i$-th feature at time $t$. Then we only change the corresponding feature vector value $X_i$ at time $t+\delta t$ if it has been changed by comparing it to the previous saved $X_i$ at time $t$. If it has been changed, only then we take it into the account to compute $\sum_{\forall i} X_i*\theta_i$ by simple addition and subtraction instead of full scale matrix multiplication.

\begin{algorithm}
 $X_n$ : saved feature vector values for $n$ comments from before for all features\;
 $\delta n$ : new comments to be processed\;
 $X^{i}_{n}$ : feature vector value of $i$-th feature for $n$ comments\;
 $X^{i}_{\delta n}$ : feature vector value of $i$-th feature for $\delta n$ comments\;
 $|X_n|$: number of total features\; 
 \ForAll{$i$ in $1,2,...,|X_n|$}{
	$X^{i}_{n + \delta n} : X^{i}_{n} + $Compute($X^{i}_{\delta n}$)\;
	}
 \caption{IncrementalFeatureExtraction()}
 \label{algo:incremental_feature_extraction}
 \vspace{-1mm}
\end{algorithm}

\subsection{Evaluation of Incremental Classifier}

We evaluated the performance of our incremental classifier component as follows, using the $983$ labeled media sessions from Vine \cite{rahatVineASONAM2015}.  We defined a baseline solution as consisting of non-incremental feature extraction and a non-incremental logistic regression classifier.  As new comments arrive for the baseline solution, it would need to recompute all feature vectors from scratch, and recompute the entire logistic regression from scratch.  We compared the total running time of the baseline solution with an incremental solution that implemented both incremental feature extraction and incremental logistic regression.  We note first that our measurements showed that the fraction of time taken by the logistic regression compared with the feature extraction time was negligible, so that total running time was dominated by feature extraction.

Figure \ref{fig:average_time_per_media_session} shows the average time taken for the standard and the incremental classifiers as the number of comments increases in media sessions.  To simplify the plot, we group the comments in sets of 10.  The time taken by the standard classification solution goes up almost linearly with the number of comments in the media session, since the standard solution must recompute all features and regression weights. On the other hand, the time taken for the incremental classifier is basically constant every time a set of 10 additional comments come in because it only has to compute the feature values for the additional 10 new comments. The justification for using 10 comments is given in section \ref{sec:priority}.

\section{Dynamic Priority Scheduler}
\label{sec:priority}
We now present the design of our dynamic priority scheduler and compare its performance with a round-robin scheduler for media sessions with no priority
assignments.
The intuition behind our scheduler is to enable our cyberbullying detection solution to process those media sessions that are likely to lead to cyberbullying
incidents earlier than the other media sessions. This would potentially
result in raising alerts much earlier, thus improving the responsiveness of
the solution. A key challenge here is how we determine the likelihood
of a media session leading to cyberbullying at a point in time. Indeed, this 
likelihood will change over time as newer comments are processed.
Thus, it makes sense to assign dynamic priority to the media sessions based on 
the current state.

Given a set of initial priorities, it is natural to consider a solution in which classification is devoted only to the high priority media session, while the lower priority cases are ignored.  Such a solution would achieve high scalability, and we term it a static priority solution.  Unfortunately, our investigation of the performance of this static priority scheduler over $983$ media sessions found the precision and recall values to be $70\%$ and $58\%$ respectively, where the recall is clearly lower than the $71\%$ recall reported in Section \ref{sec:alert} for the dynamic priority scheduler. This shows that by not considering the media sessions that are assigned low priority, we miss a significant number of actual cyberbullying sessions.  As a result, it is important to retain all sessions, as some may evolve into cyberbullying sessions despite the initial low priority prediction.  Our dynamic priority scheduler retains all sessions to improve recall, but rearranges the order in which they are processed via dynamic prioritization in order to improve responsiveness.

We now turn our attention to how to dynamically change a media session's priority after each classification by the incremental classifier component. Recall that the incremental classifier provides a confidence probability of how likely a media session contains cyberbullying. We make use of the history of these confidence values to assign a dynamic priority. The reason for using history as 
opposed to just the most recent confidence value has to do with the definition of cyberbullying.  
Cyberbullying is defined as an aggressive online behavior
that is {\it carried out repeatedly} against a person who {\it cannot easily
defend himself or herself}, creating a power
imbalance~\cite{kowalski2012cyberbullying}. To identify
repeated aggressive behavior or whether a victim can defend himself or
herself, we need to consider a much longer history than just the most recent
confidence value. We calculate the average of all past confidence values for past classifications and current classification of a particular  media session and  and compare that with a threshold value. If the average confidence 
value is more than the
threshold value, we assign a high priority to the session and if the average
value is lower than the threshold value, we assign a low priority.  Algorithm \ref{algo:setting_priority} illustrates our priority setting algorithm using an average confidence threshold ($0.2$ in this example).  Hence, we need to determine what threshold is appropriate for our solution, as explained below.

\begin{figure}
\vspace{-2mm}
\centering
\includegraphics[width=0.4\textwidth, height=0.25\textwidth]{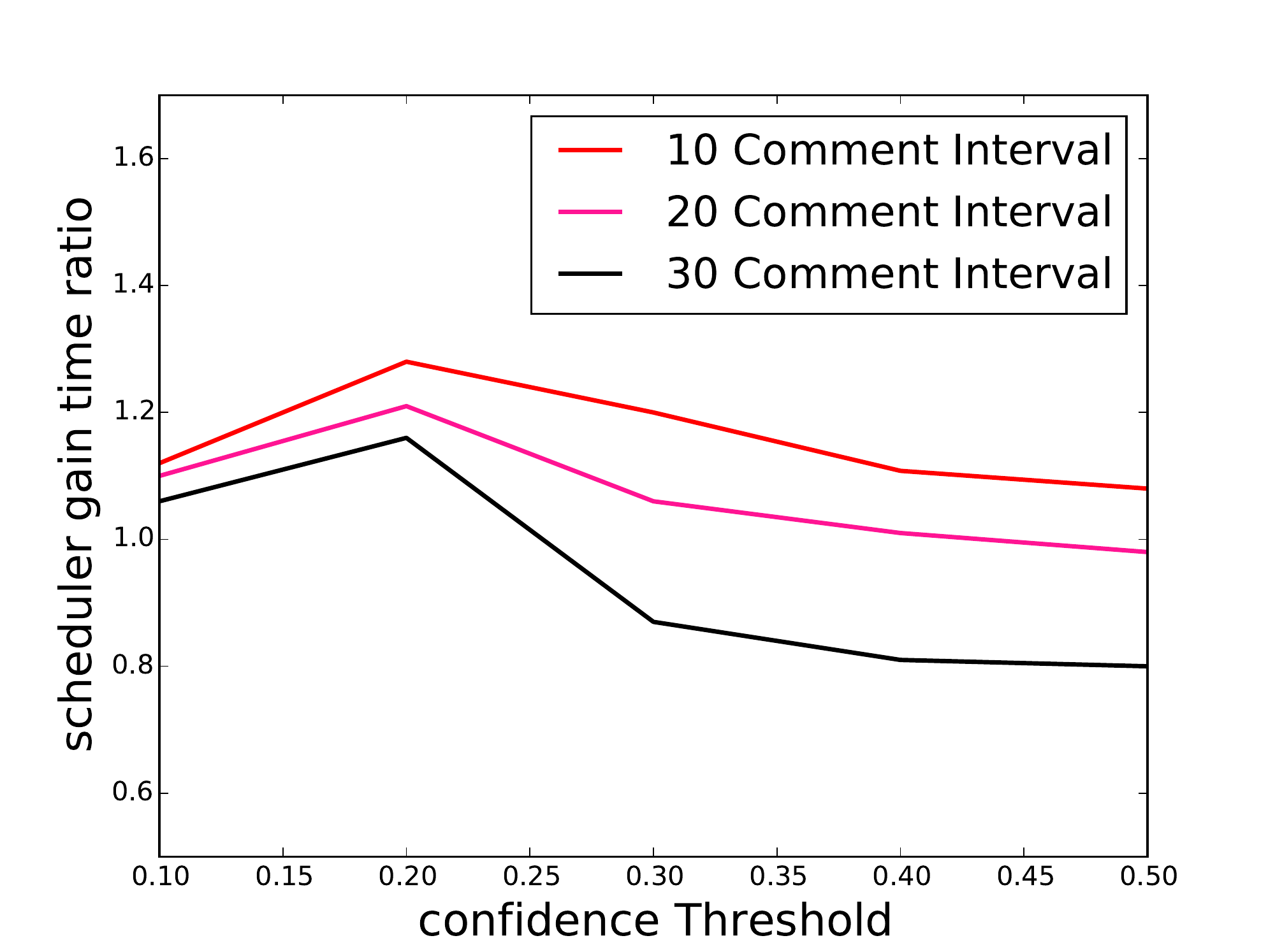}
\vspace{-5mm}
\caption{Scheduler gain time ratio for different confidence thresholds by using different comment increment sizes. }
\vspace{-5mm}
\label{fig:scheduler_gain_confidence_interval}
\vspace{-1mm}
\end{figure}

We also need to determine with what granularity our classifier ingests batches of new comments, because this affects the time to first alert.
The scheduler will choose a high priority media session to pass to the classifier.  In the time between classification attempts, a media session may receive N new comments.  If all N comments are input to the classifier at once, and N is quite large, we may delay recognizing cyberbullying, i.e., a burst of negative comments may be swamped by the other positive comments.  Therefore, we need to consider comments in small enough batches or intervals so that the classifier can catch cyberbullying with finer granularity and raise the alert early.

Figure \ref{fig:scheduler_gain_confidence_interval} assesses which combination of threshold and interval size produced the best improvement in response time using dynamic prioritization compared with a simple round-robin policy.  The round-robin scheduler is defined as one where media sessions are not assigned any priority, and the scheduler simply rotates through all media sessions, with no particular attention being paid to likely cyberbullying sessions.

As can be seen from the figure, by using a confidence threshold of $0.2$ and comment increment size of $10$, we were able to gain the maximum responsiveness over the round-robin scheduler on the $983$ Vine media sessions. We think this is because as the comment increment size goes up to $20$ or $30$, the burst of cyberbullying comments can get nullified by the other positive comments, which in turn influences the features (i.e., summation of individual comment sentiments) that are used by our incremental classifier. So $10$ comment increment size tends to be the optimal size for having enough context of a comment thread to make a knowledgeable decision about cyberbullying while also not being too big to risk being nullified by other positive comments. The table also confirms that the confidence threshold of $0.2$ offers the best speedup for our dynamic priority scheduler. For example, if a media session $m$ has been classified $3$ times at $t_1,t_2,t_3$ with classification decisions not-bullying,not-bullying,not-bullying respectively with confidence values of $0.85,0,85,0.55$, this means even though it has been classified as not cyberbullying, the confidence values of cyberbullying decision is also increasing ($0.15,0.15,0.45$) which makes it a potential candidate for a future cyberbullying session. So we take the average of the previous classification confidence values of cyberbullying class ($0.25$ in this case) and see that the average confidence value is more than $0.2$ and change the priority of this media session as high and insert it in the dynamic priority scheduler.

\begin{algorithm}
\SetAlgoLined
\ForAll{media session $m$}{
 	$Conf^{m}_{i}$: confidence value of the $i$-th comment session prediction for this media session\;
 	$n$: number of total comment session prediction in the confidence history\;
 	$Avg_{confidence}^{m}$ = $ \frac{\sum_{i=1}^{n} Conf^{m}_{i}}{n} $ \;
	
	\If{$Avg_{confidence}^{m} \geq 0.2$ and current priority is $LOW$}{
		set current Priority to $HIGH$\;
		continue\;
		}
	\If{$Avg_{confidence}^{m} < 0.2$ and current priority is $HIGH$}{
		set current Priority to $LOW$\;
		continue\;
		}
 	
 	}
\caption{SettingPriority()}
 \label{algo:setting_priority}
 \vspace{-1mm}
\end{algorithm}

\section{Alert Performance}
\label{sec:alert}

Since each media session will be passed sporadically to the classifier by the scheduler, then the classifier will generate a sequence of cyberbullying detection decisions over time for each media session.  It is therefore worth considering to what extent we should  utilize the history of detection decisions in generating the alert.  The default is to generate an alert immediately after the classifier decides that the current batch of 10 comments, in combination with earlier content, constitutes cyberbullying.  However, we wish to be sure and avoid false positives.  One option is to decouple the alert from the classification, and delay the alert until N positive decisions have been recently seen.  This design gives us some flexibility in trading off responsiveness and precision.

\begin{figure}
\vspace{-2mm}
\centering
\includegraphics[width=0.4\textwidth, height=0.25\textwidth]{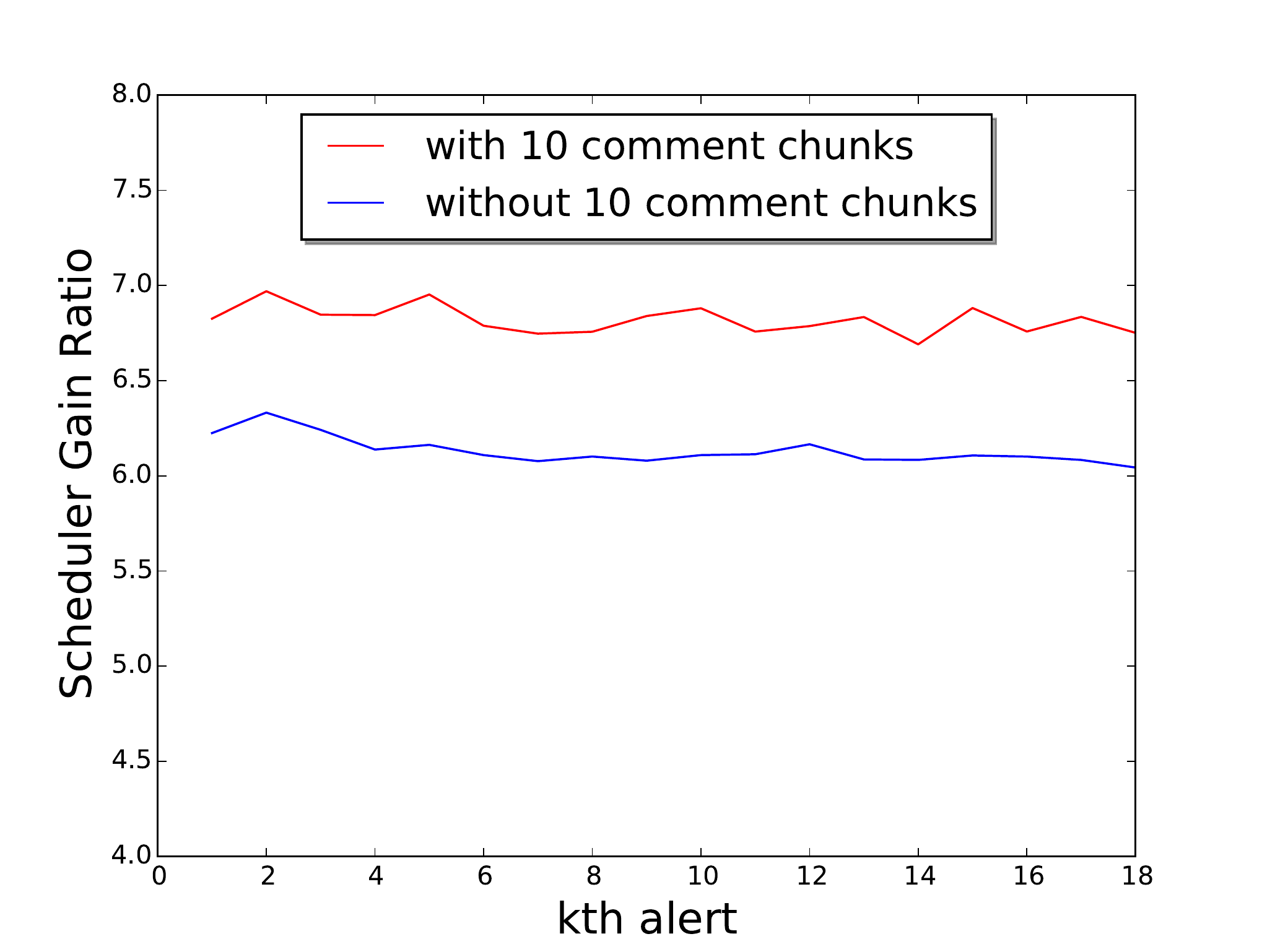}
\vspace{-5mm}
\caption{average scheduler gain time ratio for k-th alert when monitoring 100000 users}
\vspace{-5mm}
\label{fig:monitoring_scheduler_gain_time}
\end{figure}

For each media session, we maintain an array storing the results of each
classification result of that session along with the time of that classification.
We use this array to decide when to raise an alert.
In particular, we set a threshold value, which is the number of times a media session has been classified as cyberbullying since the last time an alert was raised for that session, or from the beginning if no alert has yet been raised. After experimenting
with different number of threshold values, we find that by raising an alert only when we have at least $2$ decisions for cyberbullying since the last time an alert was raised, we achieve the best precision and recall of 0.66 and 0.71 respectively, thus reducing the number of false alarms. The recall is, in fact, an improvement over the standard classifier's 0.66 (See Table \ref{table:baseline_classifier_compare_table} for comparison). 

\section{End-to-End Evaluation}
\begin{table*}
\centering
\caption{Total Time Comparison for Different Approaches and Different Number of Media Sessions (seconds)}
\label{table:final_classifier_evaluation}
\begin{tabular}{|p{5cm}||c|c|c|c|c|c|c|} \hline
Approach                                                  & 100 & 500  & 1000 & 5000       & 10000 & 50000 & 100000 \\ \hline \hline 
Standard AdaBoost                                          & 517 & 3753 & 5674 & 26784      & -     & -     & -       \\ \hline
Standard Logistic Regression                               & 104 & 526  & 1110 & 5320 (5X)  & 10438 & -     & -       \\ \hline
Incremental Classifier & 2   & 10   & 22   & 120 (223X) & 206   & 1252  & 2434     \\ \hline
\end{tabular}
\end{table*}

All our evaluations so far have been limited to the $983$ labeled media
sessions. To truly understand the extent of scalability and responsiveness 
of our solution, it is important to run a much larger set of media sessions.
To do so, we now perform our evaluations on $100,000$ Vine media sessions \cite{rahatVineASONAM2015}. For evaluating the incremental classifier's scaling performance, we compare three types of approaches, namely the best reported Vine cyberbullying classifier \cite{rahatVineASONAM2015} (Standard AdaBoost), Logistic regression without 
incremental feature extraction or classification (Standard Logistic regression), and Logistic regression with incremental
feature extraction and classification (Incremental Classifier).
Table \ref{table:final_classifier_evaluation} shows the time needed in seconds for these three approaches to process different numbers of media sessions.
First, we notice that after 5000 media sessions, AdaBoost quickly becomes unmanageable in terms of delay, exceeding the time span of a day, after which we terminated the experiment.
Next, we observe that for complete classification of $5000$ media sessions, our incremental classifier is $223$ times faster than the AdaBoost approach, which was the best classifier identified in \cite{rahatVineASONAM2015}.
Further, to classify all $100,000$ media sessions, our incremental classifier took only about $40$ minutes. 
This means that we can process at an average rate of 2,500 media sessions per minute. Note that this was achieved using a single 64-bit Windows machine 
with core i5 @1.80GHz CPU, 6GB RAM, and the solution is implemented in Python. 
We feel this makes our cyberbullying detection solution highly scalable. 

\begin{figure*}
\begin{tabular}{ll}
\hspace{-18mm}
\includegraphics[width=0.75\textwidth, height=0.3\textwidth]{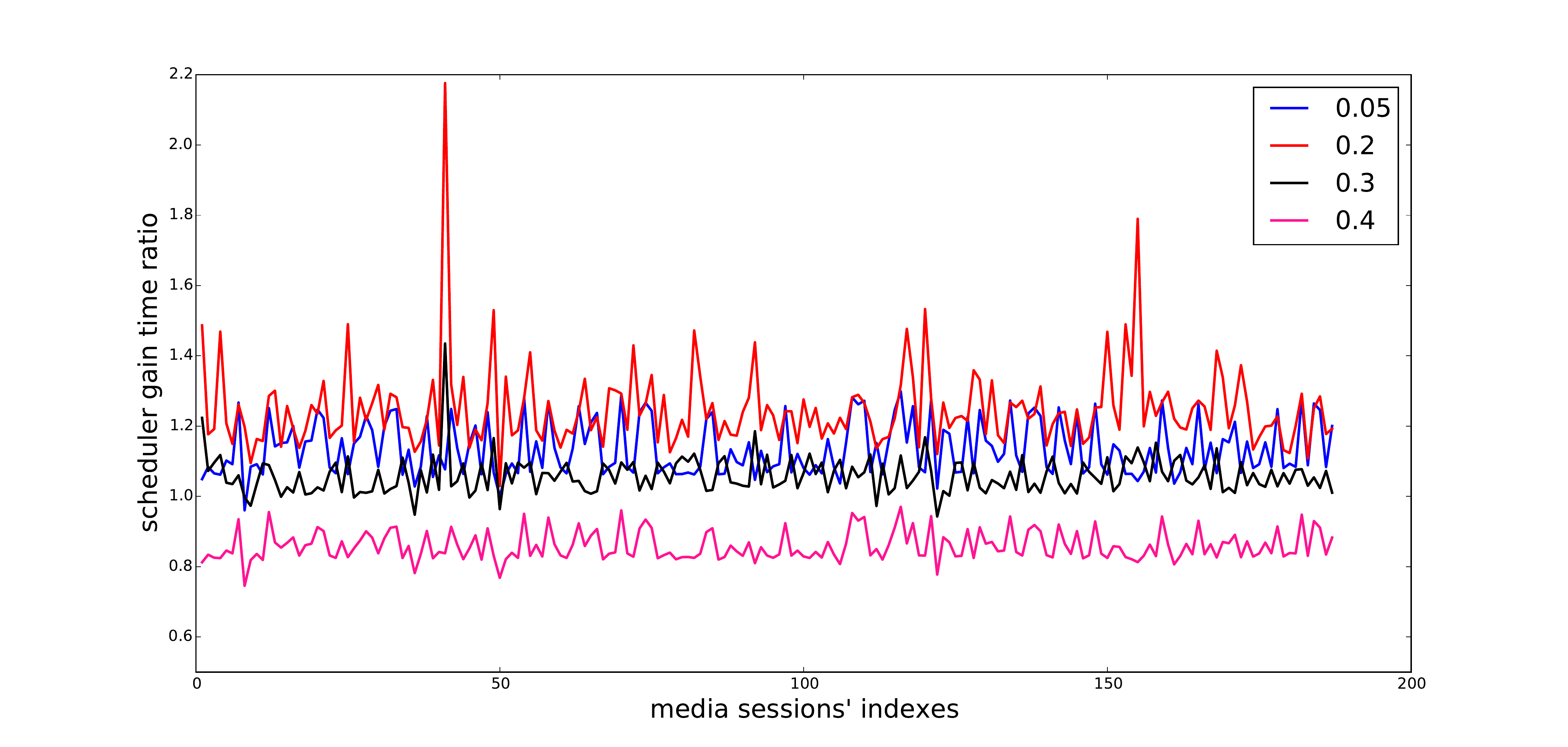}  &
\hspace{-18mm}
\includegraphics[width=0.4\textwidth, height=0.28\textwidth]{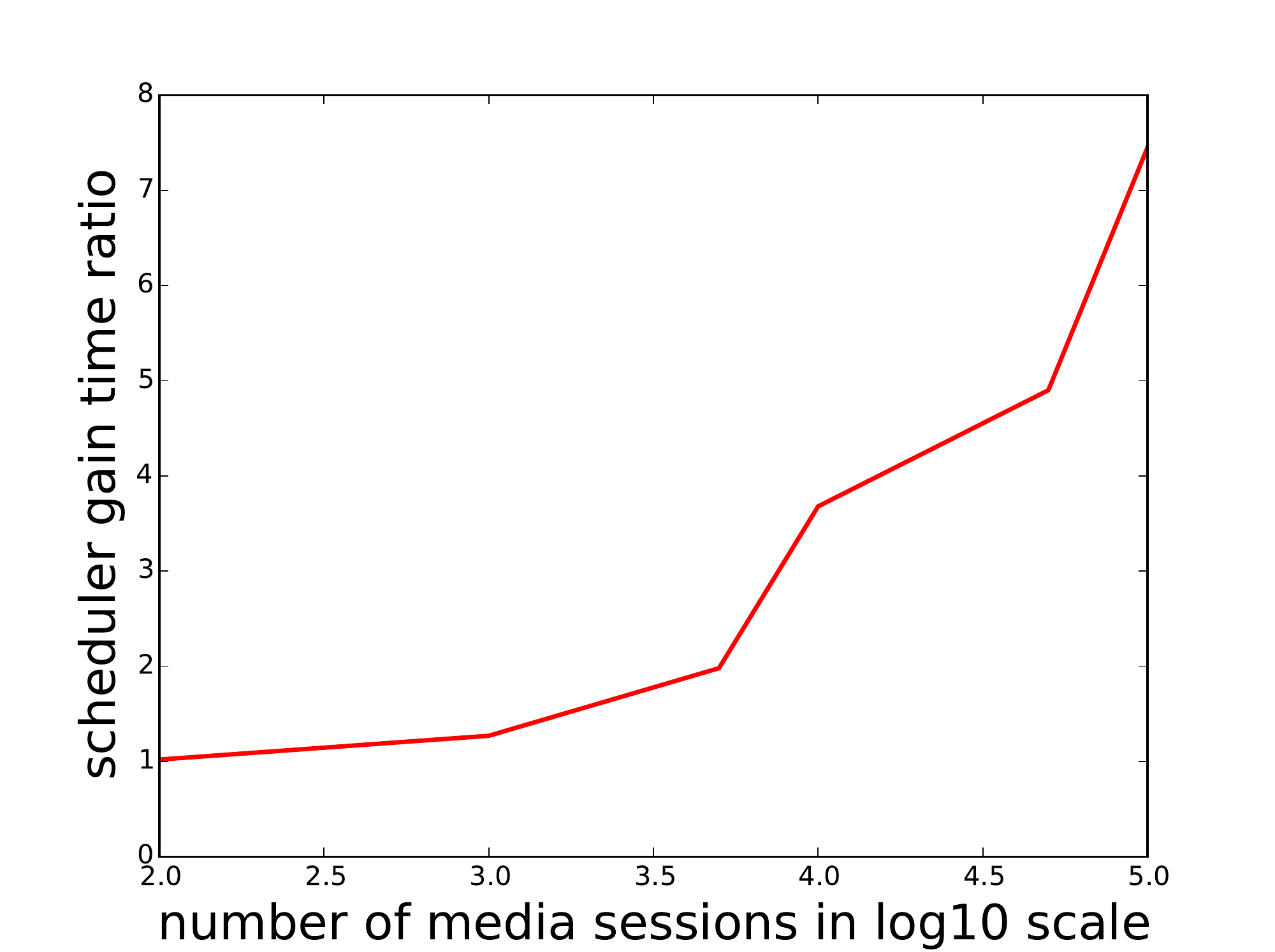} \\
\end{tabular}
\vspace{-1.0em}
\caption{(left) Scheduler gain time ratio for different confidence thresholds (right)Gain time ratio (round-robin scheduler/dynamic priority scheduler)  for different number of media sessions. }
\vspace{-1.0em}
\label{fig:scheduler_confidence_gain_combined}
\end{figure*}

Next, we investigate how our dynamic priority scheduler improves the responsiveness of our cyberbullying detection solution when compared with  a 
standard round-robin scheduler with no priority assignments for a large number of media sessions. The metrics we use to compare these two approaches is responsiveness gain, meaning the ratio of time taken by the round-robin 
schedule over the time
taken by our dynamic priority scheduler to raise an alert. 
Figure \ref{fig:scheduler_confidence_gain_combined} shows the responsiveness gain for different number of media sessions when our proposed dynamic priority scheduler is in action. As it can be seen, the gain actually tends to increase as the number of media sessions goes up, reaching almost $7$ times faster responsiveness for $100,000$ media sessions. This is because as the number of media sessions increases, so does the
number of cyberbullying media sessions. When we use our dynamic priority 
scheduler, these cyberbullying media sessions tend to get processed first as
they receive higher priority. As their numbers increase, the time savings 
due to the fact that they are processed first accumulates.
Once again, this demonstrates that our solution remains highly responsive
in the face of a large number of media sessions. On the other hand, for round-robin scheduler, because of the absence of priority, all $100000$ media sessions are processed at each pass which takes much longer than the priority scheduler where only a subset of the media sessions are processed at each step, excluding the worst case scenario. 

While the previous results were aggregated, Figure \ref{fig:scheduler_confidence_gain_combined} displays the gain times for individual cyberbullying media sessions, from the labeled set of $983$ media sessions \cite{rahatVineASONAM2015}. The findings are clearly consistent with the findings from Figure \ref{fig:scheduler_gain_confidence_interval},  which show that the biggest scheduler gain time is achieved by using a confidence threshold of $0.2$.

Next, we deploy our solution in a real world scenario to investigate the efficiency and efficacy. First, to investigate the performance of our classifier, we monitor $874$ users for $5$ days and collect the media sessions that have been classified as cyberbullying by our solution. In total $33$ media sessions were classified as cyberbullying during that period. We then employ two graduate students to manually look into those media sessions to check how many of those media sessions were actual instances of cyberbullying. After the labeling and employing the majority voting method, $24$ out of $33$ media sessions were deemed as correct classifications of cyberbullying, giving an alert accuracy of almost $73$ percent. Second, to evaluate our solution's actual efficiency in a real world scenario, we deploy our solution to monitor $100000$ users over a period of $1$ week. To compare our solution's performances, we monitor these users with two implementations, one with our dynamic priority scheduler and one with naive round robin scheduler. Figure \ref{fig:monitoring_scheduler_gain_time} shows the average gain time ratio for raising the k-th alert. We investigate the design choice of 10 comment chunks that we made in section \ref{sec:priority}. For example, if at time $t_1$, the media session is scheduled to be classified and has $15$ new comments, the first case will take only the first 10 comments and make a decision whereas the second case will take all the $15$ new comments. The way our solution is designed, a media session can generate multiple alerts during its life time. We try to compare the average scheduler gain time ratio needed for sending the k-th alert for a cyberbullying media session for the aforementioned two cases. It can be seen from figure \ref{fig:monitoring_scheduler_gain_time} that when it comes to sending alerts, on average for each media session, our dynamic priority scheduler is almost $6$ times faster than a naive round robin scheduler and by using the $10$ comment chunks design choice at each iteration of a media session's classification makes the dynamic scheduler almost $7$ times faster. \emph{This result further justifies the design choice of investigating $10$ comment chunks at each iteration for a media session that we made in section \ref{sec:priority} }.

\begin{figure}
\vspace{-2mm}
\centering
\includegraphics[width=0.4\textwidth, height=0.25\textwidth]{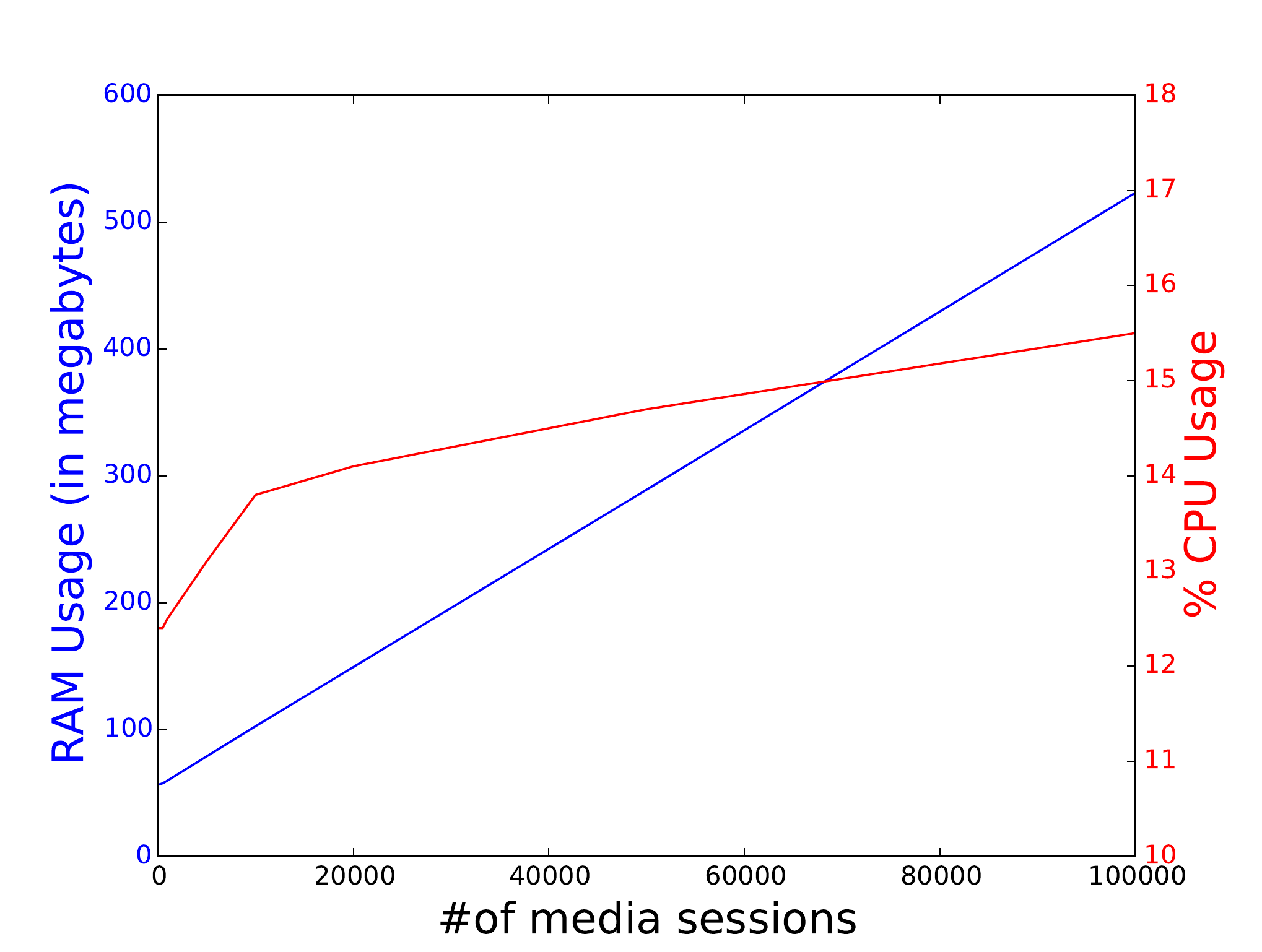}
\vspace{-5mm}
\caption{Memory and CPU Percentage usage for processing different number of media sessions}
\vspace{-5mm}
\label{fig:cpu-ram_usage}
\vspace{-1mm}
\end{figure}

Finally, we investigate the CPU and memory (RAM) usage of our solution as more and more media sessions are fed to it. We perform this experiment to check whether our proposed solution is also scalable when it comes to computational resources. Figure \ref{fig:cpu-ram_usage} shows the resource usage of our solution for different number of media sessions. We implemented our solution in a machine that had 6GB Ram available with an Intel Core i5 1.80 GHz processor. As it can be seen, as the number of media sessions go up, the memory usage increases linearly but reasonably because we tend to store the classification data in the RAM. For $100,000$ media sessions, the amount of memory and CPU percentage used is around $500$ megabytes and $15.5$ percent which is a reasonable performance. \emph{To put these performances into the perspective of real world, with a HP DL700 series server with 512GB of memory, one instance of our solution will be able to support $1.1*10^8$ active Vine media sessions where active means media sessions for which new comments are coming in. Given the fact that Vine has $8233$ media sessions shared per minute, that means we will have to add a new server with our solution installed every 8 days}. This is, of course, if we plan to keep all our media sessions, which does not have to be the case because we can filter away inactive media sessions every week for which comments are not coming in because without comments there's a very little chance of cyberbullying. By keeping in mind that in Vine, on a few percentage of total number of media sessions shared by the users contain profanity and/or potential cyberbullying instance \cite{rahatVineASONAM2015}, we can certainly keep the number of servers needed at a manageable figure by filtering away media sessions that are not potentially cyberbullying and only keep monitoring those media sessions with our server who are. We plan to delve more into this investigation in our future research work.

\section{Conclusion and Future Work}
\label{sec:conclusion} 

In this work, we have developed a multi-stage cyberbullying detection 
for online social networks, which can achieve high scalability and high responsiveness while ensuring good precision and recall. 
We have proposed three innovative techniques to accomplish this: 
(1) an initial predictor that makes fast high/low priority decisions with high 
recall for newly-created media sessions; 
(2) a dynamic priority scheduler that varies the frequency of cyberbullying 
classification for different media sessions based on the confidence of previous  
classification results; and (3) an incremental classifier to further speed up the cyberbullying detection process by reusing 
previously computed results. Using real-world data collected from the Vine 
online social network, we have demonstrated the utility of each component 
and the overall solution improvement over state-of-the-art approaches. 

As part of future work, we propose to investigate our classifier more to improve its performance, precision and recall. We also plan to build a mobile monitoring app for concerned guardians to help them get an alert whenever potential cyberbullying takes place. We also hope to expand the coverage of public social networks that our solution supports so as to monitor cyberbullying instances on a more diverse collection of social network platforms.

\bibliographystyle{abbrv}
\bibliography{rahat}  

\end{document}